\documentclass{article}
\usepackage{spconf,amsmath,graphicx}
\usepackage{mathtools}
\usepackage{eufrak}
\usepackage{algorithm}
\usepackage{array}
\usepackage{booktabs}
\usepackage{algorithmicx}
\usepackage{algpseudocode}
\usepackage{multirow}
\usepackage{blindtext}
\usepackage{stfloats}

\title{A Robust Speaker Clustering Method Based on Discrete Tied Variational Autoencoder}
\name{Chen Feng, Jianzong Wang\sthanks{Corresponding author: Jianzong Wang, jzwang@188.com}, Tongxu Li, Junqing Peng, Jing Xiao}
\address{Ping An Technology (Shenzhen) Co., Ltd.}
\begin{document}
%
\maketitle
\begin{abstract}
Recently, the speaker clustering model based on aggregation hierarchy cluster (AHC) is a common method to solve two main problems: no preset category number clustering and fix category number clustering. In general, model takes features like i-vectors as input of probability and linear discriminant analysis model (PLDA) aims to form the distance matric in long voice application scenario, and then clustering results are obtained through the clustering model. However, traditional speaker clustering method based on AHC has the shortcomings of long-time running and remains sensitive to environment noise. In this paper, we propose a novel speaker clustering method based on Mutual Information (MI) and a non-linear model with discrete variable, which under the enlightenment of Tied Variational Autoencoder (TVAE), to enhance the robustness against noise. The proposed method named Discrete Tied Variational Autoencoder (DTVAE) which shortens the elapsed time substantially. With experience results, it outperforms the general model and yields a relative Accuracy (ACC) improvement and significant time reduction.
\end{abstract}
\section{INTRODUCTION}
\noindent
Due to rapidly growing demand for organizing a mast of recorded speech, speaker clustering methods have been developed quickly and used widely. Speaker clustering is an effective means to reduce the task of massive voice management. At the moment, there are two main application scenarios. One is the speaker diarization task. The goal is to detect speaker segments and each unique speaker will be identified by a single label \cite{8600111} \cite{1677976} \cite{8461666}. Under normal conditions, segments that need to be clustered are mostly short. The other is the speaker linking task for speaker embedding \cite{cumani2019exact} \cite{sturim2016speaker} \cite{khoury2014hierarchical}. This paper concerns the application of second scenarios. Firstly, it converts utterances to a unique speaker vector, and this paper investigates the speaker clustering method based on i-vectors. Secondly, utterances of each class will be from a single speaker by clustering model. In many cases, there has no idea that how many classes are going to cluster and the utterances are mostly long. For speaker linking task, AHC model becomes the preferred clustering method \cite{Solomonoff1998Clustering} without knowing the number of default categories.

The original speaker clustering method based on AHC has been popular in speaker linking task. When the quality of utterances is better, the result is more satisfactory. This method consists of several steps: Firstly, training a feature extractor like based on Gaussian mixture model (GMM)and universal background model (UBM) \cite{Dunn2000Approaches} to extract i-vectors \cite{Dehak2011Front}. Secondly, training a PLDA \cite{Ioffe2006Probabilistic} model to score similarity between different utterances. Then the PLDA scorings are normalized (p-scoring) in the range between 0 and 1. Finally, $1-$p-scoring is used to express the distance between corresponding utterances in AHC model. Clustering process stops until meeting an appropriate threshold or getting a required categories number. Note that PLDA is a linear Gaussian generation model. When distribution of features is close to the Gaussian distribution, the strong practicability scorings make original AHC method works well.

Time-consuming is another disadvantage of the speaker clustering method based on original AHC. Before clustering, it is necessary to get standardized similarity scorings among all utterances. In the case of large amount of data, the calculation burden of this part is very heavy. A common strategy is to put similar utterances into the same groups through mapping models before clustering. Then AHC model is used in each groups, and the utterances between different groups no longer perform clustering operations, which means p-scorings between different groups do not need to be calculated. In recent years, many speaker clustering methods have been proposed successively, which relying on other models before AHC model to get the initial clustering results. For example, phone clustering \cite{Shen2011Speaker} can be applied firstly to construct the universal phone cluster models. Combining speaker QBE like Hash method \cite{sturim2016speaker} with canopy cluster model also is a better way to make up the shortfall of traditional method.
\begin{figure*}[htb]
\centering
\includegraphics[height= 3.2in,width=6.8in]{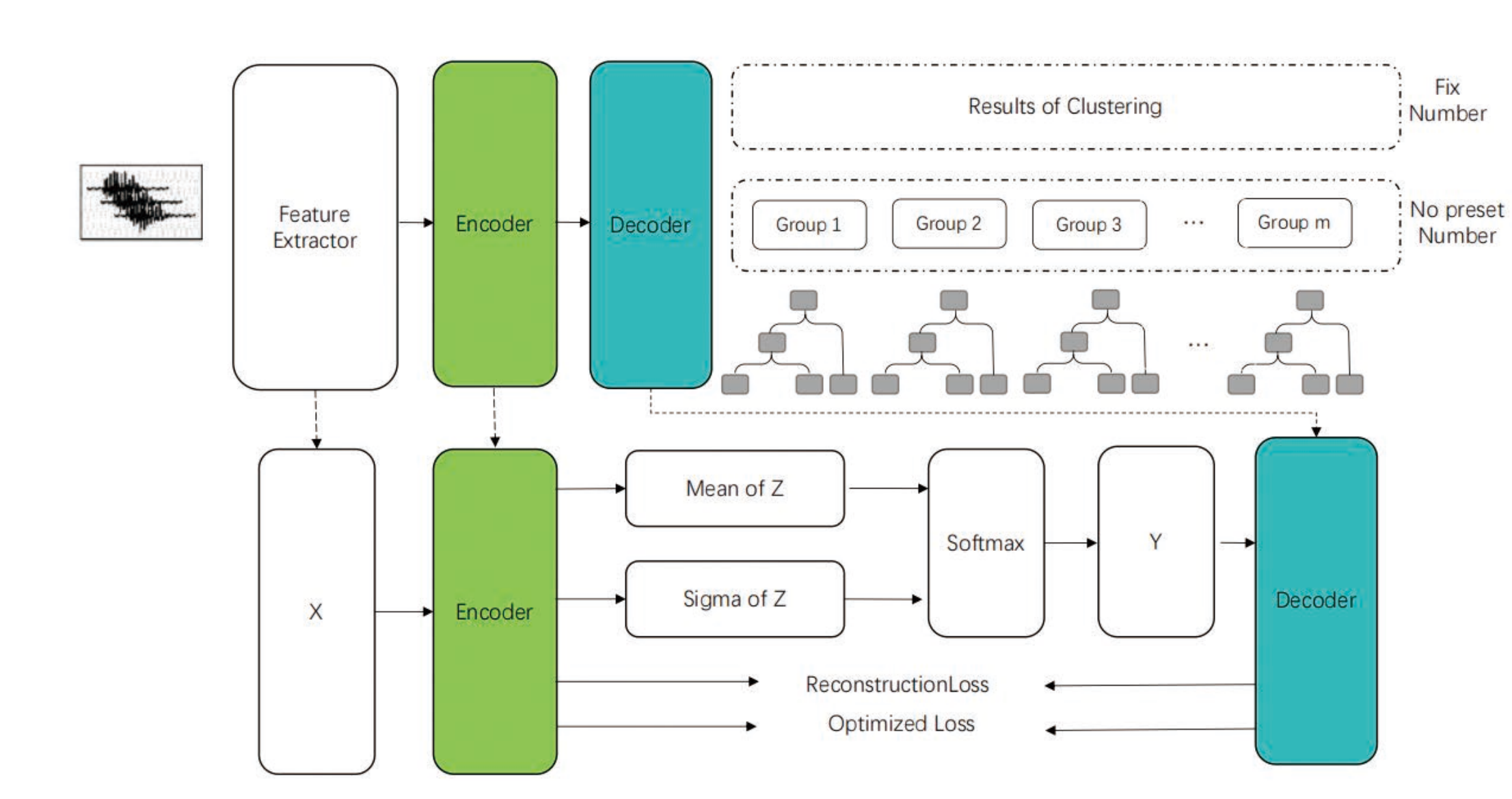}
\caption{The Structure of Speaker Clustering Method DTVAE}
\label{fig_sim}
\end{figure*}

More recently, variational autoencoder model has achieved incredible performance in speaker recognition area. By transforming vectors into approximate Gaussian distribution \cite{Zhang2019VAE}, the latent vectors of the variational autoencoder are more suitable for PLDA scorings. As the extension of the original variational autoencoder \cite{Kingma2013Auto} formulation, TVAE \cite{villalba2017tied} model performs non-linear discriminant analysis well which also can be regard as the replacement of PLDA model. Variational autoencoder also has shown great potential in another area of clustering tasks. Variational Deep Embedding \cite{Jiang2016Variational} is a unsupervised clustering method outperforms other clustering models on 5 benchmarks. To complete the fundamental clustering task in image recognition \cite{dilokthanakul2016deep}, it can study a variant of the variational autoencoder model with a Gaussian mixture as a prior distribution to solve over-regularization.

The motivation for this paper is making up the weaknesses about time-consuming and exploring more useful methods that can learn more complex  distributions of features. Moreover, these methods are robust to the environment noise. Inspired by TVAE, it is first applied to complete initial speaker clustering task as a mapping model. Not only does it work well to accomplish speaker clustering task, but also avoids unstable results suffering from random number in hash method. Recently witnessing some remarkable attempts to learn by maximizing Mutual Information in variational autoencoder model or dual autoencoder network\cite{yang2019deep}, and the learned representations are more robust to noise. Moreover, Mutual Information \cite{kraskov2005hierarchical} \cite{Slonim2000Document} estimation enhances discriminative information between inputs from different classes. Adding Mutual Information \cite{yang2019deep} to the framework of variational autoencoder model still works well in the image clustering tasks. As such, we add the Mutual Information to our robust speaker clustering method. The contribution of the implementation is as follows:

\begin{itemize}
  \item[$\bullet$] The method proposed in the paper is an innovative application of variational autoencoder in the field of speaker clustering. It Converts the continuous variable representing the speaker's information to the discrete variable as the classes information, that is adding a softmax layer in the structure of model.
  \item[$\bullet$] The method proposed in the paper makes the features closer to the real distribution and get the better result of clustering combined with the consideration of MI.
  \item[$\bullet$] When clustering with fix category number, the method save the time for training a scoring model and the time for getting the scorings. When clustering with no preset category number, a better initial grouping is given, so the time for calculating scorings between different initial groups has been saved.
\end{itemize}

\section{SPEAKER CLUSTERING METHOD}

\noindent
As stated in Figure 1, the proposed clustering method DTVAE shown in upper part of figure which can complete two main clustering issues. When there has the fix category number, it can get the result of clustering directly. When there has the no preset category number, it gives initial groups. With enlightenment of TVAE model, converting the speaker factor variables as the discrete variables to stand for different classes in DTVAE. It is aimed to map similar features into the same groups, and eliminate the need to think about clustering between different groups. Then just need to cluster within the same group through AHC model.

Meanwhile, in the training process, not only the maximization of the lower bound of evidence considered in DTVAE, but also the optimization idea of Mutual Information was added. Adding Mutual Information in the loss function to make the model more robust. The lower part of the figure shows AHC models which are employed in the divided groups, then by combining results of each groups, we can form the final clustering result.

The framework of DTVAE model is stated in lower part of Figure 1, where speaker factor $Y$ as classes variable. Therefore, and $Y$ is also a discrete variable. To assume that $Y=\{y_1,y_2,...y_m\}$ obeys multinomial distribution, and limit distribution of which is a normal distribution. Similarly, framework of DTVAE model divided into encoder and decoder. In encoder, it defines $X_i=\{x_{ij}\}^{N_{i}}_{j=1}$ and $Z_i = \{z_{ij}\}^{N_{i}}_{j=1}$ for representing i-vector that belong to the $i$ speaker and latent vector, where $N(i)$ is the number of utterances for speaker $i$. The encoder takes $x_{ij}$ as input and computes the mean and convariance of the latent vector $z_{ij}$ together with classed vector $y_i$ though softmax layer as output. 

\begin{equation}\label{eq1}\
\begin{split}
q_{\phi}(y_i|X_i) = &\Omega(mt_i,mt_i(1-t_i))\sim N(y_i|\mu,\Sigma')\\
\mu = &(I+\sum^{N_i}_{j=1}(\Sigma_{\phi}(x_{ij}^{-1}-I))^{-1}\\
\Sigma' = & \Sigma_{\phi}(X_i)\sum^{N_i}_{j=1}\Sigma_{\phi}(x_{ij})^{-1}\mu_{\phi}(x_{ij}))\\
\end{split}
\end{equation}
In decoder, it takes $y_i$ and $z_{ij}$ as input and output mean and covariance of the conditional likelihood of $x_{ij}$. By hypothesis, $x_{ij}$ also obeys a conditional normal distribution. 

To learn parameters of DTVAE, it optimizes variational Bayes lower bound (ELBO), which is equivalent to maximize marginal likelihood.
\begin{equation}\label{eq1}\
\begin{split}
log P_{\theta}(X_i) \geq L(X_i,\theta,\phi) &= E(log\frac{P_{\theta}(X_i,y_i,z_i)}{q_{\phi}(y_i,z_i|X_i)})\\
max \sum^m_{i=1}logP(X_i|\theta) &\Rightarrow max \sum^m_{i=1}L(X_i,\theta,\phi)\\
\end{split}
\end{equation}
It's similar to the TVAE model \cite{villalba2017tied}, the reconstruction loss function that needs to maximize can be defined as:
\begin{equation}\label{eq1}\
\begin{split}
L_r = &-\sum^m_{i=1}L(X_i,\theta,\phi) \approx \sum ^m_{i=1} \{KL(q_{\phi}(y_i|X_i)||p_{\theta}(y_i))\\
+\sum^{N_i}_{j=1}&(KL(q_{\phi}(z_{ij}|y_i,x_{ij})||p_{\theta}(z_{ij}))-E(log p_{\theta}(x_{ij}|y_i,z_{ij})))\}.
\end{split}
\end{equation}
Under the posteriors, employing reparametrization trick \cite{rezende2014stochastic} to get the latent vector.

Nevertheless, the above loss function $L_r$ cannot well guarantee the quality of speech information expressed by the vector. Inspired by Mutual Information, adding it into the DTVAE model to obtain better result. In the process of training, the larger the Mutual Information is, the more effectively the correlations are measured among i-vectors $X$, latent variable $Z$ and classes factor $Y$. Considering the defect of unbounded divergence of KL divergence, Jensen-Shannon divergence \cite{liese2006divergences} is adopted to replace it. The optimized loss \cite{yang2019deep} function that needs to minimize can be defined as:
\begin{equation}\label{eq1}
\begin{split}
L_j = &-\beta(E_{p_{\theta}}[log(\sigma(D(x_{ij},y_i,z_{ij})))]\\
+&E_{q_{\phi}}[log(1-\sigma(D(x_{ij},y_i,z_{ij})))]).
\end{split}
\end{equation}
where $\sigma(.)$ is a discriminator, where $\sigma(x)= sigmoid(x)$. The concrete form of $D(.)$ defined as
\begin{equation}\label{eq1}
D(x_{ij},y_i,z_{ij})=log\frac{2q_{\phi}(z_{ij},y_i|x_{ij})}{q_{\phi}(z_{ij},y_i|x_{ij})+p_{\theta}(x_{ij}|y_i,z_{ij})}.
\end{equation}
To avoid duplication, it has omit the same part as reconstruction loss function.

Considering all the items, the definition of total loss of DTVAE model is
\begin{equation}\label{eq1}
\begin{split}
min _{\theta,\phi }L_z =min _{\theta,\phi } ( L_r+L_j )
\end{split}
\end{equation}

\section{EXPERIMENT}
\subsection{Experiment Setting}
\noindent
Voxceleb1 is a public data set. The utterances are from real English speech completely, and the number of utterances is 145265 from 1251 speakers. The data set is balanced in gender, and what it is important that the utterances contain real noise rather than artificial white noise, which includes environmental noise, sound recording equipment noise, laughter, echo, and etc. For the speaker linking task with long utterances, it samples 43733 utterances as dataset from it and real time all greater than 9s.

To evaluate clustering result, here adopts Accuracy (ACC) as standard evaluation index in this paper. In dataset $X=\{x_1,...x_n\}$, all utterances from $P=\{s_1,...,s_m\}$ speakers are clustered into $C=\{c_1,...c_p\}$. 
The ACC calculation used is
\begin{equation}\label{eq1}
ACC = max_b\frac{\sum^n_{i=1} 1(t_i=b(c_i))}{n},
\end{equation}
where $t_i$ and $b(c_i)$ are true label and predicted cluster of point $x_i$.

By \emph{Kaldi}, we train a GMM-UBM model of 2048 components and i-vectors extractor of dimension 600 by the train set. The DTVAE method consists of 4 layers. They are input layer, hidden layer, output layer and softmax layer,where \emph{relu} and \emph{tanh} \cite{Katz2017Reluplex} as the non-linear activations. This paper focuses on improving shortcomings of traditional speaker clustering method, as such here set the original method (i-vectors-PLDA-AHC) as the baseline model here.

\subsection{Experiment Results}
\noindent
When meeting the issue with the no preset category number, Table 1 presents the ACC for our novel speaker clustering method and the baseline model in the case of mass data with a large number of classes. Here set the number of initial classes is 3 and random choose different amount of utterances from dataset. As more utterances that belongs to the same speaker, DTVAE may performs better than baseline model. As the amount of data increases, ACC index decreases to some extent. But sometimes may be a small improvement. It is difficult to improve performance while reducing time consumption. In detail, through normality test \cite{Shapiro1965An} i-vectors do not obey the Gaussian distribution which will make the original AHC model error too large to use ineffective, which is the main reason for original AHC having poor performance. When the volume of utterances belonging to the same speaker increases, the advantages of our method model will be more prominent.

In term of time consumption reduction, DTVAE takes less time to run which save the time of calculating the scorings between different groups. As shown in Figure 2, the blue and red line show time required of ours and baseline, and the orange line express percentage of reduced time of diverse datasize. It can significantly reduce the time required as the amount of data increases. With the number of utterances is 10000, our method yields a significant 32\% time reduction. In the case of more utterances, the time consumption is reduced more.

When meeting the issue with the fix category number, we also analyze its potential, which completes the clustering task alone. In the case of a small number of classes. DTVAE model may get the higher accuracy. Table 2 presents the ACC for ours model and baseline model with each speaker has an average of 50 utterances, which come from the mean of 3 randomly experiences results. Here is one point to emphasize, Generally, baseline clustering method needs a long time to train PLDA model. Moreover our method omits this process, and only relying on the utterances to be clustered to train the parameters of the network structure has the chance to get a good result. That point also provides an innovative idea for the traditional speaker clustering study.
\begin{table}[!htbp]

    \centering
    \fontsize{9}{12}\selectfont    
    \caption{ACC(\%) for proposed  method and baseline model}
    \begin{tabular}{cccc}
    \toprule
    \multirow{2}{*}{Data Size}&
    \multicolumn{2}{c}{ACC} &\cr
    \cmidrule(lr){2-4}
    &  DTVAE & Baseline  \cr
    \hline
    \hline
      1000  & 70.1\% & 68.6 \% \cr
      3000 & 68.9\% &64.4\% \cr
      5000 & 62.2\% &57.1\%&  \cr
      10000 & 55.6\% &52.2\%&  \cr
    \bottomrule
    \end{tabular}\vspace{0cm}
    \label{tab:Training_sizes}
\end{table}
%
\begin{table}
    \centering
    \fontsize{9}{12}\selectfont    
    \caption{ACC(\%) for DTVAE model and baseline model}
    \begin{tabular}{ccccccc}
    \toprule
     \multirow{2}{*}{Classes Number}& &
    \multicolumn{2}{c}{ACC} &\multicolumn{2}{c}{} \cr
    \cmidrule(lr){2-6}
    & & DTVAE & Baseline & & \cr
    \hline
    \hline
      3  &\vline& $\mathbf{93.3}\%$ & 86.6\%  &  \cr
      4  &\vline& 80.8\% & 85.5\%   & \cr
      5  &\vline& 74.0\% & 78.0\% &\cr
      6  &\vline& $\mathbf{83.3}\%$& 83.3\%   &\cr
      7  &\vline& 85.7\% & 89.2\%   & \cr
      8  &\vline& $\mathbf{88.8}\%$ & 85.0\%   & \cr
    \bottomrule
    \end{tabular}\vspace{0cm}
    \label{tab:Training_sizes}
\end{table}

\begin{figure}[!t]
\centering
\includegraphics[height= 2.3in,width=3.4in]{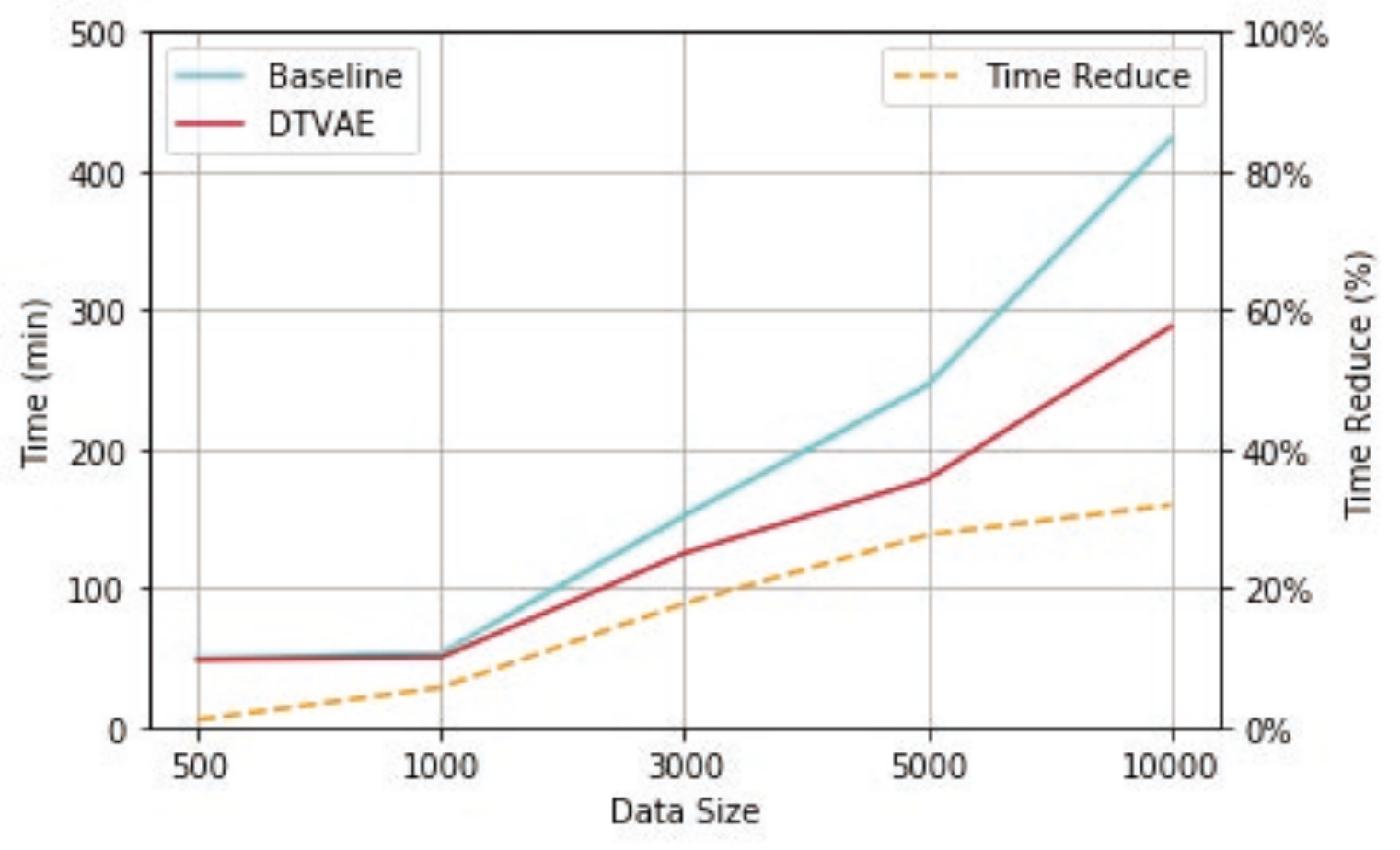}
\caption{The Percentage of Reduced Time}
\label{fig_sim}
\end{figure}

\section{CONCLUSION}
\noindent
This paper proposes a robust speaker clustering method based on Tied Variational Autoencoder and Mutual Information. It not only reduces the time consumption of the whole clustering process, but also is robust to environment noise. First, discrete Tied Variational Autoencoder, which regarded as the extension of TVAE model, is utilized to get the initial clustering groups with no preset category number. Furthermore, AHC model is employed in all groups individually. In term of fix category number issue, better results are given diretly by DTVAE. The experimental evaluation shows that our method outperforms the baseline algorithm (i-vectors-PLDA-AHC) in terms of ACC and time. This is only a preliminary work, and there is still much to be investigated. In the future, hopefully, it can complete speaker clustering task alone by deformation of DTVAE model.
\section{ACKNOWLEDGMENTS}
\noindent
This paper is supported by National Key Research and Development Program of China under grant No.2018YFB1003500, No.2018YFB0204400 and No.2017YFB1401202. \\
Corresponding author is Jianzong Wang from Ping An Technology (Shenzhen) Co., Ltd.
\bibliographystyle{IEEEbib}

\clearpage

\end{document}